\documentclass{icrc2009}

\usepackage{graphicx}   
\usepackage[caption=false]{caption}    
\usepackage[font=footnotesize]{subfig} 
\usepackage{fixltx2e}
\usepackage{url}

\newcommand{\shorttitle}[1]%
{\markboth{Proceedings of the 31\MakeLowercase{$^{st}$} ICRC, {\L}\'{o}d\'{z} 2009}{#1} }
\newcommand{\etal}{\MakeLowercase{\textit{et al. }}} 

\begin{document}
\title{Acoustic detection of high energy neutrinos in ice: Status and results from the South Pole Acoustic Test Setup}

\author{\IEEEauthorblockN{Freija Descamps\IEEEauthorrefmark{1} for the IceCube Collaboration\IEEEauthorrefmark{2}}
                            \\
\IEEEauthorblockA{\IEEEauthorrefmark{1}Department of Subatomic and Radiation Physics, University of Ghent, 9000 Ghent, Belgium,\\
	          \IEEEauthorrefmark{2}See the special section of these proceedings.}}
\shorttitle{Freija Descamps \etal SPATS status and results }
\maketitle

\begin{figure*}[th]
  \centering
  \includegraphics[width=5.2in, height=4.1in]{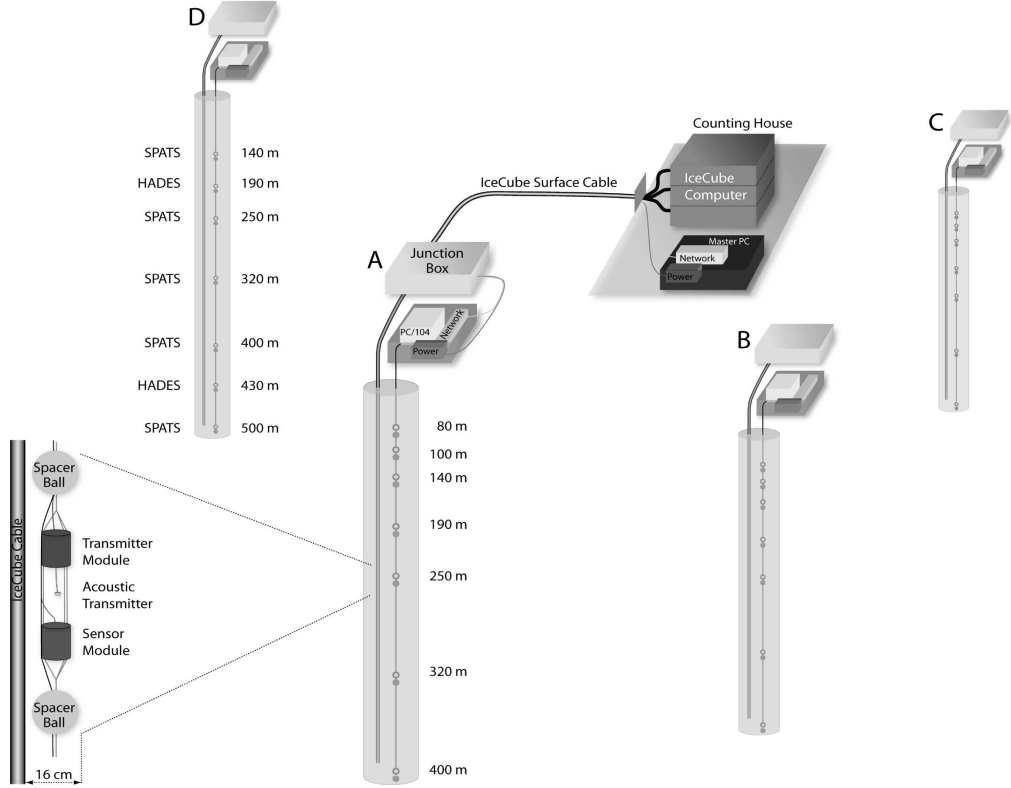}
  \caption{Schematic of the SPATS detector.}
 \label{fig:1}	
 \end{figure*}
\begin{abstract}

The feasibility and specific design of an acoustic neutrino detection array at the South Pole depend on the acoustic properties of the ice. The South Pole Acoustic Test Setup (SPATS) has been built to evaluate the acoustic characteristics of the ice in the 1 to 100~kHz frequency range. 
The most recent results of SPATS are presented.
\end{abstract}

 
\section{Introduction}
The predicted ultra-high energy (UHE) neutrino fluxes from both hadronic processes in cosmic sources and interaction of high energy cosmic rays with the cosmic microwave background radiation are very low. Therefore extremely large detector volumes, on the order of 100~km$^{3}$ or more, are needed to detect a significant amount of these neutrinos.
The idea of a large hybrid optical-radio-acoustic neutrino detector has been described and simulated in~\cite{hybrid1,hybrid2}.
The density of detectors in such a possible future UHE neutrino telescope is dictated by the signal to noise behaviour with travelled distance of the produced signals. The ice has optical attenuation lengths on the order of 100~m, enabling the construction and operation of the IceCube~\cite{Achterberg:2006md} detector.
The relatively short optical attenuation length makes building a detector much larger than IceCube prohibitively expensive. In contrast, the attenuation lengths of both radio and acoustic waves are expected to be larger~\cite{barwick,price} and may make construction of a larger detector feasible.
The South Pole Acoustic Test Setup (SPATS) has been built and deployed to evaluate the acoustic attenuation length, background noise level, transient rates and sound speed in the South Pole ice-cap in the 1 to 100~kHz region so that the feasibility and specific design of an acoustic neutrino detection array could be assessed. 
Acoustic waves are bent toward regions of lower propagation speed and the sound speed vertical profile dictates the refraction index and the resulting radius of curvature. An ultra-high energy neutrino interaction produces an acoustic emission disk that will be deformed more for larger sound speed gradients and so the direction reconstruction will be more difficult. A good event vertex reconstruction allows accurate rejection of transient background. The absolute level and spectral shape of the continuous background noise determine the threshold at which neutrino induced signals can be extracted from the background and thus set the lower energy threshold for a given detector configuration. Transient acoustic noise sources can be misidentified as possible neutrino candidates, therefore a study of transient sources and signal properties needs to be performed. The acoustic signal undergoes a geometric 1/r attenuation and on top of that scattering and absorption. The overall acoustic attenuation influences detector design and hence cost.
\section{Instrumentation}
\subsection{The SPATS array}
The South Pole Acoustic Test Setup consists of four vertical strings that were deployed in the upper 500 meters of selected IceCube holes~\cite{Achterberg:2006md} to form a trapezoidal array, with inter-string distances from 125 to 543~m. Each string has 7 acoustic stages. 
Figure~\ref{fig:1} shows a schematic of the SPATS array and its in-ice and on-ice components.
It also shows a schematic drawing of an acoustic stage comprised of a transmitter and sensor module. 
The transmitter module consists of a steel pressure vessel that houses a high-voltage pulse generator board and a temperature or pressure sensor. 
Triggered HV pulses are sent to the transmitter, a ring-shaped piezo-ceramic element that is cast in epoxy for electrical insulation and positioned $\sim$13~cm below the steel housing. Azimuthally isotropic emission is the motivation for the use of ring shaped piezo-ceramics. The actual emission directivity of such an element was measured in azimuthal and polar directions~\cite{fisher}.
The sensor module has three channels, each 120$^{\circ}$ apart in azimuth, to ensure good angular coverage.  
\subsection{The retrievable pinger}
A retrievable pinger was deployed in 10 water-filled IceCube holes down to a depth of 500~m: 6 holes were pinged in December 2007-January 2008 and 4 more holes were pinged using an improved pinger design in December 2008-January 2009. The emitter\footnote{ITC-1001 from the International Transducer Company} is a broad band omni-directional transmitter which has a  transmission power of 149~dB~re~($\mu$Pa/V) at 1 meter distance. 
Upon receiving a trigger signal, a short ($\sim50\mu s$) HV pulse is sent to the piezo-ceramic transmitter that is suspended about 2~m below the housing. A broadband acoustic pulse is then emitted. Both the SPATS array and the pinger are GPS synchronized so that the arrival times of the pinger-pulses can be determined. Data was collected for all SPATS-sensor levels at 80, 100, 140, 190, 250, 320, 400, 430 and 500~m depth and the pinger was pulsed at repetition rates of 1, 8 or 10~Hz. At the SPATS instrumented depths, the pinger lowering was stopped for five minutes so that a signal could be recorded with every SPATS sensor at that pinger position. For the December 2008-January 2009 holes, no stop was made at 80, 100, 140 and 430~m depth.
\section{Results}
\subsection{Sound speed}
The sound speed analysis uses data from December 2007-January 2008 geometries where the pinger and sensor were at the same depth and 125~m apart. The pinger emitter was situated in a column of water where no shear waves can propagate, nevertheless shear waves were generated at the water-ice boundary where mode conversion was expected. Therefore transit times can be extracted from the data for both pressure and shear waves for many instrumented SPATS levels. There is agreement with~\cite{Weihaupt} for the pressure wave measurement in the not fully compacted ice of the upper region (firn). The extracted shear wave speed is about half of the pressure wave speed, as expected~\cite{Albert}.
A linear fit was made to the data in the deep and fully compacted ice between 250 and 500~m depth. 
We find following results for the pressure and shear wave sound speeds ($v_{p}$ and $v_{s}$) and their variation with depth (gradient: $g_{p}$ and
$g_{s}$):
\begin{eqnarray*}
 v_{p}(375m)=(3878\pm 12) m/s,\\
 g_{p}=(0.09 \pm 0.13) (m/s)/m, \\
 v_{s}(375m)=(1975.8 \pm 8.0) m/s,\\
 g_{s}=(0.067 \pm 0.806) (m/s)/m.
\end{eqnarray*} 
The gradient for both pressure and shear waves is consistent with zero. Both sound speed measurements are performed with a better than 1$\%$ precision, taking into account the errors on the horizontal distance, pinger and sensor depths, emission and arrival times. For more details on the SPATS sound speed analysis, see~\cite{freija}.
\subsection{Gaussian noise floor}
The noise is monitored in SPATS through a forced 200~kHz read-out of all sensor channels for 0.5~s every hour. The distribution of ADC counts in each of the operational channels is Gaussian and stable during the present observation time of more than 1 year with a typical deviation of the mean noise level of $\frac{\sigma_{RMS}}{<RMS>}~<~10^{-2}$. 
The SPATS sensors on strings A, B and C have been calibrated in liquid water at 0$^{\circ}$C in the 10 to 80~kHz frequency range prior to deployment~\cite{fisher}. Lab measurements have shown~\cite{fisher,timoposter} that the sensitivity of the SPATS sensor in air at atmospheric pressure increases by a factor of  1.5~$\pm$~0.2 when cooled down from 0$^{\circ}$C to -50$^{\circ}$C. We use this value to estimate the sensitivity of the SPATS sensors after they were deployed in the cold Antarctic ice. A measurement of the sensitivity at room temperature as a function of static pressure was performed in a pressure vessel and the results indicate a change in sensitivity of $<30\%$ in the 1 to 100 bar region~\cite{timoposter}.
The noise level and fluctuations are high for all 4 strings in the firn region, where the transition from a snow/air mixture to compact bulk ice takes place. This is consistent with the effect of a large sound speed gradient in that layer so that the surface noise is refracted back to the surface. In the fully compacted ice, below the firn, noise conditions are more stable and we can derive an average noise level below 10mPa in the relevant frequency range (10 to 50kHz). For more details on the SPATS noise floor analysis, see~\cite{timo}.
 \begin{table*}[th]
  \caption{SPATS attenuation length studies.}
  \label{table_att}
  \centering
  \begin{tabular}{|c|c|c|c|c|}
  \hline
   Attenuation analysis  & $\lambda$ (m) & uncertainty & comment on uncertainty \\
   \hline 
    Pinger energy time domain& 320 & 60m & standard deviation of distribution \\
    Pinger energy frequency domain & 270& 90m & standard deviation of distribution \\
    Interstring energy all levels & 320& 100m & standard error of weighted mean of distribution\\
    Interstring energy 3-level ratio & 193 & -&best fit \\
  \hline
  \end{tabular}
  \end{table*}
\subsection{Transient events}
\begin{figure}[h]
 \centering
 \includegraphics[width=7.0cm,height=6.0cm]{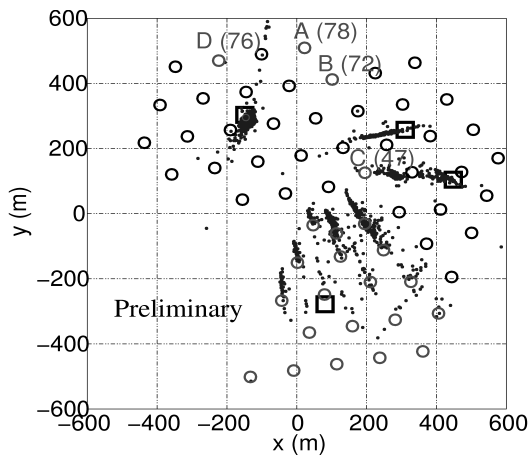}
 \caption{An overview of the spatial distribution of transient events as detected by SPATS between September 2008 and April 2009. The circles indicate the positions of the IceCube holes, the SPATS strings are indicated by their corresponding letter-ID. The ``Rodriguez well'' are represented by squares. The smearing-effect is an artefact of the event reconstruction algorithm due to not accounting for refraction in the firn.}
\label{transients}
\end{figure}

The SPATS detector has been operated in transient mode for 45 minutes of every hour since August 2008. If the number of ADC counts on any of the twelve monitored channels exceeds a certain level above noise, we record a 5ms window of data around the trigger on that channel. 
The resulting trigger rate is stable and on the order of a few triggers every minute for each of the twelve monitored channels. Most of these events are Gaussian noise events, where only one sample is outside the trigger boundaries. The transient events are processed off-line and analysed for time-coincidence clustering. Figure~\ref{transients} shows the spatial distribution of a total of 4235 reconstructed transient events as detected by SPATS between 1 September 2008 and 23 April 2009 (4422 hours integrated lifetime). The data shows clear and steady sources or 'hot spots' that can be correlated with the refreezing process of the water in sub-surface caverns.
 Each IceCube drilling season, a ``Rodriguez wel'' (RW) is used as a water reservoir. The main source of transients is the 07-08~RW. There is also steady detection of the 05-06/04-05~RW. The 06/07~RW was a steady source until October 2008. The 08-09~RW has not yet been detected. Transient data-taking continued during the IceCube 08-09 drilling season and the refreezing of 12 holes nearest to the SPATS array are audible whereas 7 of the farthest are not. No vertices were reconstructed deeper than 400~m. 
\subsection{Attenuation length}
We present here 2 classes of attenuation analyses, all of which use the sensors on the  frozen-in SPATS strings but with a different sound source: the retrievable pinger and the frozen-in SPATS transmitters. Table~\ref{table_att} gives an overview of the SPATS attenuation length studies with their respective uncertainties.
\begin{figure}[h]
 \centering
 \includegraphics[width=7.5cm,height=4.5cm]{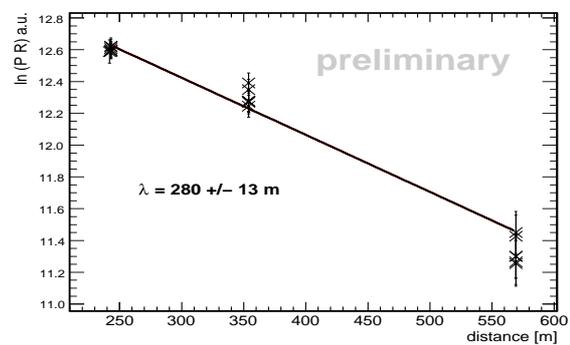}
 \caption{An example of a (ln(amplitude$\cdot$distance) vs. distance)-fit for a single channel at 400~m depth.}
\label{singlefit}
\end{figure}
\subsubsection{Pinger attenuation length}
\begin{figure}[h]
 \centering
 \includegraphics[width=7.5cm,height=5.5cm]{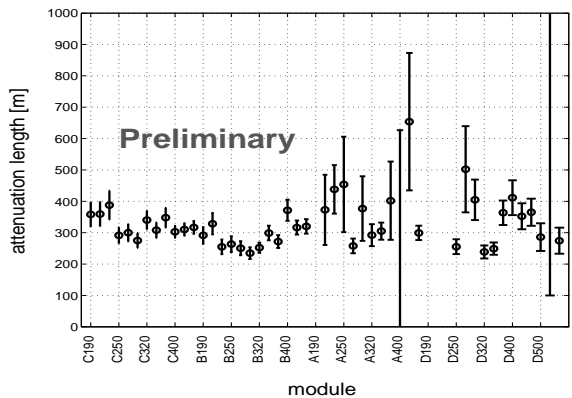}
 \caption{Pinger energy analysis result from time domain for all recorded levels.}
\label{att-dist}
\end{figure}

The retrievable pinger was recorded simultaneously by all SPATS sensors for each of the 4 IceCube holes in which it was deployed during the December 2008-January 2009 period. The pinger holes were almost perfectly aligned relative to the SPATS array, making the single-channel analysis independent of polar and azimuthal sensitivity variation. The energy of the signal, which dominates in the frequency range from 5 to 35~kHz, was extracted both from the waveforms in the time domain (4 pinger holes) and from the power spectrum (3 pinger holes). Figure~\ref{singlefit} shows an example of a single-level fit for the energy in the frequency domain (single channel at 400~m depth).  For both analyses, the quoted value in Table~\ref{table_att} is the mean of all fits with the standard deviation as error. The result for the time domain analysis uses all levels as shown in Fig.~\ref{att-dist} and points which have large error bar ($\frac{\lambda}{\sigma_{\lambda}}<3$) have been excluded (4/47 combinations). There is no evidence for a depth-dependence of the attenuation length.
\subsubsection{Inter-string attenuation length}
The complete SPATS inter-string set of April 2009 consists of all possible transmitter-sensor combinations excluding the shallowest transmitters. The transmitters were fired at 25Hz repetition rate and for each combination a total of 500 pulses was recorded simultaneously by the three channels of the sensor module. An averaged waveform can therefore be obtained and the pressure wave energies extracted. An averaged noise-waveform allows us to subtract the noise contribution.
Every transmitter in SPATS can be detected by 3 sensors at the same depth and different distances (strings), this means that only the unknown sensor sensitivities have to be included into the systematic error for a single-level attenuation length fit. We have fitted all data of the transmitters that are detected at three different distances. Each single fit does not constrain the attenuation length very well due to the large sensor-to-sensor variations in sensitivity. By combining all fits, we find a weighted mean and the standard deviation of the weighted mean as value for the attenuation length and its error (Table~\ref{table_att}).
A way to work around the missing calibration of the sensors and transmitters is to build ratios of amplitudes using two transmitter-sensor pairs. For isotropic sensors and transmitters, one such measurement should yield the attenuation length. To minimize the remaining variation of the sensitivity due to the varying polar angle, we limited the amplitude ratios to the levels at 250, 320 and 400~m depth. This ratio-method yields an attenuation length with a large systematic error due to unknown azimuthal and remaining polar sensor orientations, therefore only the best fit is quoted in Table~\ref{table_att}.
\section{Conclusions}
We have presented the most recent results from the SPATS setup. Both pressure and shear wave speeds have been mapped versus depth in firn and bulk ice. This is the first measurement of the pressure wave speed below 190~m depth in the bulk ice, and the first measurement of shear wave speed in the South Pole ice. The resulting vertical sound speed gradient for both pressure and shear waves is consistent with no refraction between 250 and 500~m depth.
Extrapolating sensitivities from laboratory calibrations gives a first estimation of the absolute noise level at depths larger than 200~m and indicates values below 10~mPa integrated over the 10 to 50~kHz frequency range. The in-ice transient rates are low, most of the transient events can be correlated with well-know anthropogenic sources. We have presented an overview of the different SPATS attenuation length studies, the preliminary results show that the analyses favour attenuation lengths in the 200-350~m region. The current dataset does not allow a distinction to be made between absorption- or scatter-dominated attenuation length and dedicated measurements are under consideration.
\section{Acknowledgment}
We are grateful for the support of the U.S. National Science Foundation and the hospitality of the NSF Amundsen-Scott South Pole Station.

\end{document}